\documentclass{jpp}

\usepackage{graphicx}
\usepackage{natbib}

\ifCUPmtlplainloaded \else
  \checkfont{eurm10}
  \iffontfound
    \IfFileExists{upmath.sty}
      {\typeout{^^JFound AMS Euler Roman fonts on the system,
                   using the 'upmath' package.^^J}%
       \usepackage{upmath}}
      {\typeout{^^JFound AMS Euler Roman fonts on the system, but you
                   dont seem to have the}%
       \typeout{'upmath' package installed. JPP.cls can take advantage
                 of these fonts, if you use 'upmath' package.^^J}%
      }
  \else
  \fi
\fi

\ifCUPmtlplainloaded \else
  \checkfont{msam10}
  \iffontfound
    \IfFileExists{amssymb.sty}
      {\typeout{^^JFound AMS Symbol fonts on the system, using the
                'amssymb' package.^^J}%
       \usepackage{amssymb}%

      }{}
  \fi
\fi

\ifCUPmtlplainloaded \else
  \IfFileExists{amsbsy.sty}
    {\typeout{^^JFound the 'amsbsy' package on the system, using it.^^J}%
     \usepackage{amsbsy}}
    {}
\fi

\newsavebox{\astrutbox}
\sbox{\astrutbox}{\rule[-5pt]{0pt}{20pt}}

\title[The Equatorial Current Sheet]{The Equatorial Current Sheet and other interesting features of the
Pulsar Magnetosphere}

\author[I. Contopoulos]%
{I\ls o\ls a\ls n\ls n\ls i\ls s\ns C\ls o\ls n\ls t\ls o\ls p\ls o\ls u\ls l\ls o\ls s$^{1,2}$%
  \thanks{Email address for correspondence: icontop@academyofathens.gr}}

\affiliation{$^1$Research Center for Astronomy and Applied Mathematics, Academy of Athens\\4 Soranou Efessiou Str., Athens 11527, Greece\\[\affilskip]
$^2$National Research Nuclear University, 31 Kashirskoe highway,
Moscow 115409, Russia}

\pubyear{2010}
\volume{650}
\pagerange{119--126}
\date{ ; revised  ; accepted  .}
\begin{document}

\maketitle

\begin{abstract}
We want to understand what drives magnetospheric 
dissipation in the equatorial current sheet.
Numerical simulations  have limitations and, unless we have a clear
a priori understanding of the physical processes involved, their
results can be misleading. We argue that the canonical pulsar
magnetosphere is strongly dissipative and that a
large fraction (up to $30-40\%$ in an aligned rotator)
of the spindown luminosity is redirected 
towards the equator where it is dissipated into particle acceleration
and emission of radiation. We show that this is due to the 
failure of the equatorial electric current to cross the
Y-point at the tip of the corotating zone.
\end{abstract}


\section{Limitations of numerical simulations}

Numerical simulations help us understand how real pulsars work.
Unfortunately, they have severe limitations: 

\subsection{Lego\textsuperscript{TM} tiles}

The radius of a neutron star is about 10km. The light cylinder
lies a few to a few thousand times away. The polar cap is a few to
a hundred times smaller. It is thus very hard to equally resolve
in 3D the stellar vicinity and the magnetosphere over several
light cylinder radii, with the possible exception of millisecond
pulsars. The current state-of-the-art 3D numerical resolution
consists of at most a few hundred grid cells in each direction inside
the light cylinder. This is manifestly insufficient to capture and
resolve certain important features of the pulsar magnetosphere that
are seen in high-resolution 2D solutions. Ten years ago, \citet{2006MNRAS}
analyzed the axisymmetric pulsar magnetosphere in the highest yet
achieved resolution of more than 10,000 grid cells inside the
light cylinder. Even though that resolution is not sufficient to
resolve the internal structure of the current sheet, it taught us
a couple of interesting things about the magnetospheric Y-point
(the tip of the corotating -so called `dead'- zone) that are hard
to see in 3D simualtions:
\begin{enumerate}
\item The poloidal magnetic and electric fields right above and
below the Y-point both drop to zero. Thus, the charge density at
the Y-point also drops to zero \citep[see also][]{2003ApJ...598..446U}.
\item In the ideal
MHD solution \citep[][hereafter CKF]{1999ApJ...511..351C},
the current sheet extends inwards of the Y-point along the
boundary (separatrix) of the dead zone. In that
region, the current sheet is charged oppositely to the charge of
the equatorial current sheet, and its charge density changes
discontinuously to zero at the Y-point. In particular, as the
Y-point approaches the light cylinder, the electric and magnetic
fields and the charge density right interior to it diverge
\citep{2003ApJ...598..446U,
2009A&A...496..495K}.
\end{enumerate}

These features of the current sheet are completely lost in current
3D numerical simulations.
What is most disconcerning, is that in order to determine the
pulsar energy losses, i.e. its spindown rate, we need to know the
position of the Y-point since this is what determines how many
field lines remain closed in the dead zone, and how many carry
electromagnetic radiation. In numerical simulations that begin
with a stationary star that is then set to uniform rotation, it
has been observed that magnetic field lines open up after a
fraction of one stellar rotation and a Y-point forms initially at
some distance $r_{\rm Y}$ {\em inside} the light cylinder (the
cylindrical radius $r_{\rm LC}\equiv c/\Omega$, where $\Omega$ is
the stellar rotational angular velocity). When that happens, the
solution settles to a perfectly valid (`happy') steady-state that
emits electromagnetic energy at a rate $L(r_{\rm Y})\approx
(r_{\rm Y}/r_{\rm LC})^{-2}L(r_{\rm LC})>L(r_{\rm LC})$
\citep[see also][]{2005A&A...442..579C, 2006MNRAS}.
The Y-point then gradually evolves towards the
final steady state where the Y-point reaches the light cylinder.
Depending on the numerical resolution, that evolution takes place
at a much longer timescale (several rotational periods) than the
timescale (fraction of one period) needed for the original happy
steady-state to establish itself. Notice that in practice, the Y-point
is determined as the innermost tip of the equatorial current sheet where
it reaches the corotating dead zone.

This may very well be a serious problem in the numerical
determination of the pulsar spindown rate. What we are really
interested to know is $L(r_{\rm LC})$ because we believe that as
the pulsar spins down, the corotating dead zone manages
to always lie close to the
instantaneous light cylinder, thus $r_{\rm Y}\approx r_{\rm LC}$. This 
is achieved via continuous reconnection at the Y-point through which
more and more formerly open field lines close and increase the
size of the dead zone.
If that is not the case, and during the small amount of time
available to run our numerical simulation the magnetosphere
has relaxed to an $r_{\rm Y}$ significantly different from $r_{\rm
LC}$ \citep[see Fig.~1; compare also to Fig.~3
of][]{2015ApJ...801L..19P}, then the
numerical estimate of the spindown luminosity $L(r_{\rm LC})$ is
unreliable. Notice that in reality, $r_{\rm Y}$ may not keep up
with the light cylinder as the pulsar spins down, and this may account
for the measured values of pulsar braking indices that are found
to be less than the canonical value of 3
\citep[see also][for an alternative view]{1998FCPh...20....1K}.
The numerical evaluation of the real pulsar braking index is, however,
way beyond the capabilities of modern computers.

\begin{figure}
\centering
\begin{minipage}{0.5\textwidth}
\centering
\includegraphics[trim={4.5cm 2cm 2.5cm 6cm}, width=\textwidth]{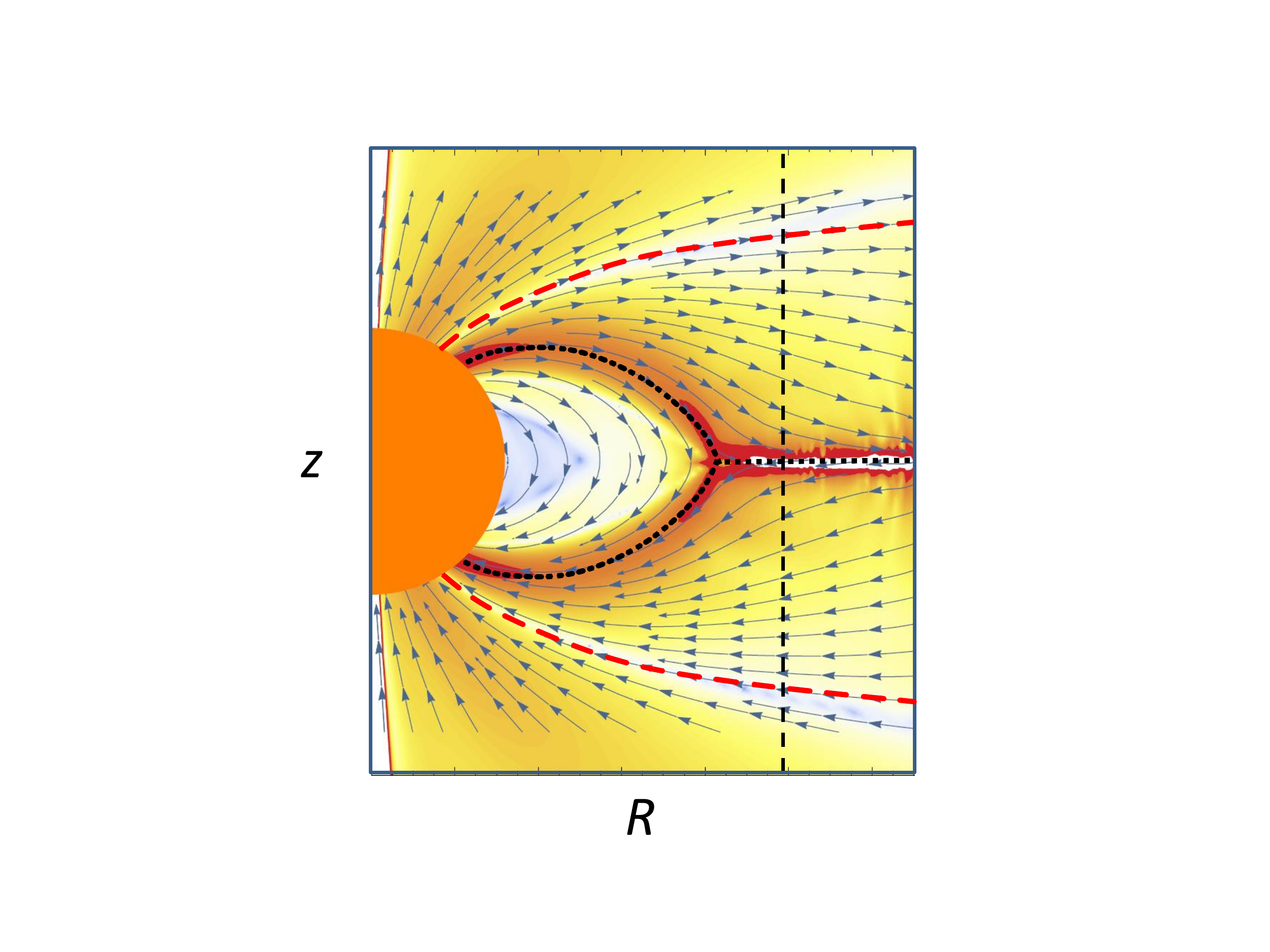}
\end{minipage}%
\begin{minipage}{0.5\textwidth}
\centering
\includegraphics[trim={2.5cm 2cm 4.5cm 6cm}, width=\textwidth]{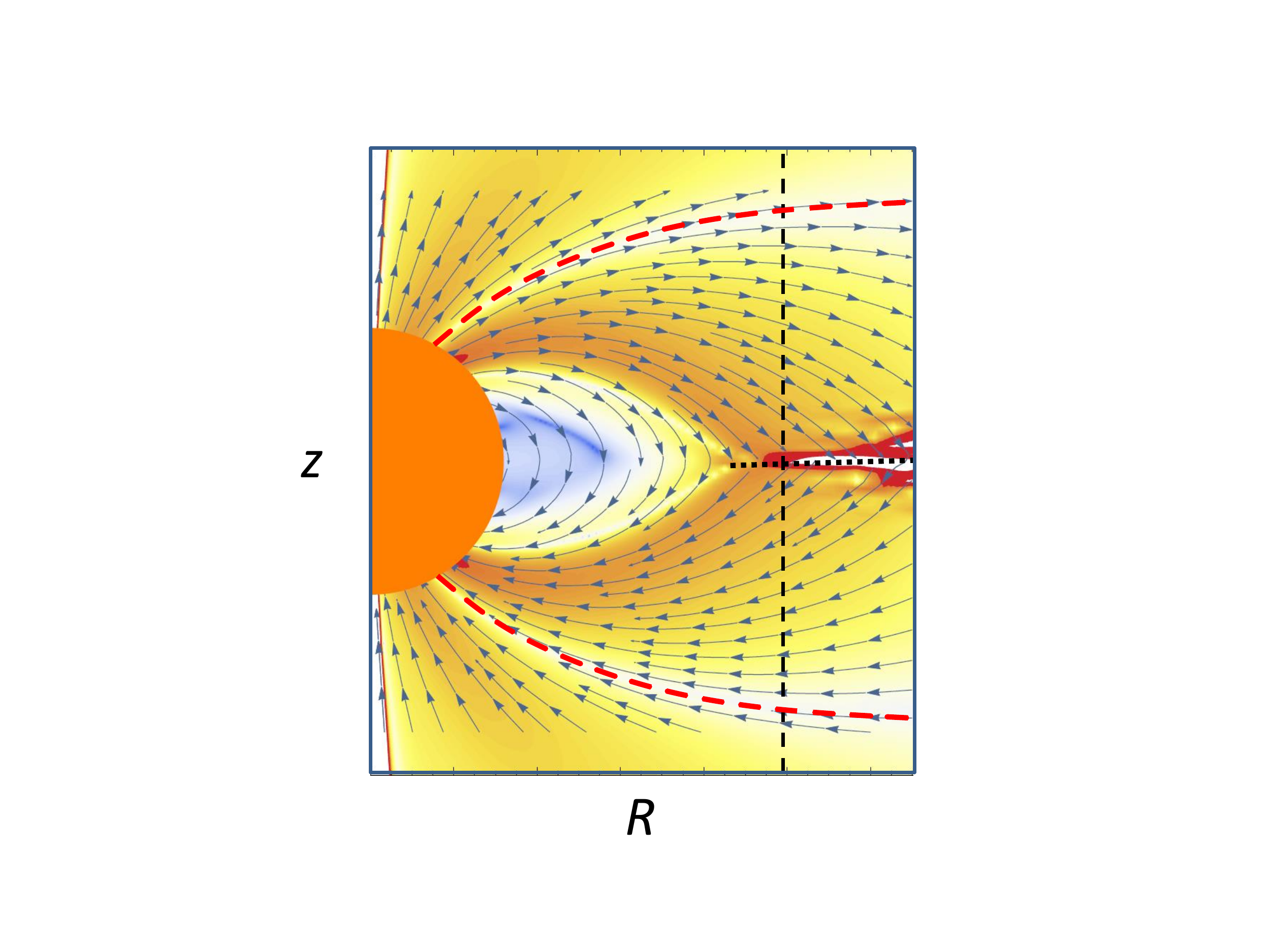}
\end{minipage}%
\caption{Examples of ideal MHD numerical simulations 
with Y-points clearly inside the light cylinder. These particular
simulations run only for 6 stellar rotations which apparently wasn't enough
time for them to relax to a final steady state.
Vertical dashed line: line cylinder. Lines with arrows: poloidal magnetic field.
Color scale: magnitude of poloidal electric current density multiplied by $R^2$
(white regions: no poloidal currents; red regions: poloidal current sheets,
also drawn with black dotted lines). The lines of zero poloidal electric current that separate
the regions of outgoing and ingoing electric currents are shown with red short
dashed lines (see also Fig.~2).
Orange semicircle: neutron star. Left:
$E$ is enforced to be less than $B$ everywhere. In practice, this implies
that current sheets are treated as contact discontinuities.
Both the equatorial and separatrix
current sheets form. The equatorial current sheet
extends clearly inside the light cylinder.
The current sheet appearing inside the Y-point
is clearly an artifact of our particular treatment of contact disctontinuities.
Right: We allow $E$ to take any value compared to $B$, and
current sheets are treated with a dissipative Aristotelian prescription \citep{2013arXiv1303.4094G}.
In that case, only the equatorial current sheet
survives, and the separatrix current sheet disappears, as
expected in \citet{2014ApJ...781...46C}, and seen also very clearly in Fig.~3 of
\citet{2015MNRAS.448..606C}.}
\label{Fig1}
\end{figure}

\subsection{Death Stars\textsuperscript{TM}}

In Hollywood fiction, Death Stars\textsuperscript{TM} shoot
intense beams of energy on their own. Real pulsars work differently.
A certain magnetospheric structure
is established that {\em demands} a certain flow of charged
particles through it in order to support the so-called
Goldreich-Julian electric charge density $\rho_{\rm GJ}$
\citep{1969ApJ...157..869G} and the
CKF electric current density $J_{\rm CKF}$ throughout
\citep[the first time this name was used was in][]{2010MNRAS.404..767C}.
There is a clear distinction here between the
so-called Goldreich-Julian electric current density defined as
$J_{\rm GJ}\equiv \rho_{\rm GJ}c$ in the overall outward direction
along magnetic field lines, and the actual electric current
density $J_{\rm CKF}$ that is required throughout the
magnetosphere in order to establish a force-free ideal MHD
equilibrium.
In general, $J_{\rm CKF}$ has nothing to do
with $J_{\rm GJ}$, except in the asymptotic region
$r\gg r_{\rm LC}$ where $J_{\rm CKF}\rightarrow J_{\rm GJ}$.
CKF were the first to argue that the establishment of
this particular electric current distribution is {\em as
important} for the global solution {\em as} the establishment of the well known Goldreich-Julian
electric charge distribution.

How the magnetosphere achieves this configuration (i.e. what is
the particular type, distribution, and kinematics of the charge
carriers that fill the magnetosphere) remains a matter of active
research and disagreement (see section 4 below). There seems
to exist several (an infinity?) ways to achieve a happy
configuration, and Nature decides which one to establish,
but one thing is certain: the star does not shoot charged particles
on its own. We are concerned about the results of numerical
simulations because whenever the
density drops below a limit that a code cannot handle (a so
called `density floor'), the code freely supplies
matter in the `problematic' regions. It is often seen that
the value of the density floor or the particle replenishment rate
affect the outcome of the numerical evolution
\citep{2015ApJ...801L..19P}. Even if such floors
do exist in nature in the form of microscopic processes that e.g.
freely generate electron-positron pairs wherever they are needed
\citep[e.g.][]{2014ApJ...795L..22C}, their values are
more than twenty orders of magnitude beyond the capabilities
of modern supercomputers
in a global magnetospheric simulation. Indeed,
the most recent calculations contain about $10^8$ particles over
a volume on the order of $r_{\rm LC}^3$ and reach sigma 
values on the order of $10^4$. Assuming a neutron star
magnetic field on the order of $10^{12}$~G, this corresponds to
particles with charge and mass more than twenty orders of
magnitude larger than the charge and mass of the electron,
and number densities more than twenty orders of magnitude
smaller than the corresponding Goldreich-Julian
densities at the light cylinder.

We  believe that global numerical simulations cannot yet teach us
what is the type of charge carrier that fills the magnetosphere,
and where are these particles generated. It is too
early for an ab initio reconstruction of the pulsar magnetosphere,
and we need to return to our drawing boards and reconsider the
global picture.

\subsection{Trade secrets}

We would like to acknowledge a danger with time dependent ideal
MHD numerical simulations in their treatment of current sheets.
A current sheet obviously lies outside the regime
of ideal MHD, so, mathematically, it should be treated as an
unresolved contact discontinuity. Obviously, numerical codes
encounter problems at the locations where the current sheet begins
to form. These problems are treated with special custom techniques
developed by each programer for his/her own code (the
`trade secrets'). In particular,
how ${\bf E}\cdot{\bf B}=0$ is enforced, and how we make sure that
$E$ nowhere exceeds $B$ (in ideal MHD) yield a particular time
dependent evolution in each particular code. The important goal of
course is for all different codes to be able to reproduce the gross
features of the final steady state obtained by CKF and described in extreme detail
by \citet{2006MNRAS}. 
Different codes yield slightly different results regarding the detailed structure
around the current sheet \citep[e.g. oppositely charged layers above and below
the equatorial current sheet seen in][, a current sheet right inside
the Y-point as seen in Fig.~1, etc.]{2010MNRAS.404..767C,
2012MNRAS.420.2793K}.
Unfortunately, there exist few high resolution test
solutions and all of them are axisymmetric and steady-state.


The message we want to convey here is that existing numerical
codes have been `tweaked' to reproduce the magnetospheric current
sheets appearing in the few known axisymmetric ideal MHD solutions
available in the literature. Currently there are no 3D tests of 3D
codes. Most importantly though, because of their particular
`tweaks', we cannot yet trust their time-dependent evolution in more
sophisticated simulations like evolving binary neutron star
magnetospheres. This may potentially be problematic in the
numerical simulations of the transient
electromagnetic signal expected when two magnetospheres colide
during the merger of two neutron stars
or two black holes \citep[e.g.][]{2013HEAD...1312101B,
2014PhRvD..90d4007P}.

\section{What drives magnetospheric dissipation?}


Let us remind the reader how we arrived at our current
understanding of the pulsar magnetosphere. The basic picture of a
unipolar inductor was formulated by \citet{1969ApJ...157..869G}
and a cartoon of the magnetospheric structure was drawn. In that
picture, electric charges fill and electric currents flow through
the magnetosphere, in such a way that an outward ${\bf E}\times
{\bf B}$ Poynting flux vector develops everywhere. Another fact
that was obvious from the very first discussion on pulsar
magnetospheres is that the magnetosphere, if ideal, must open up
beyond a distance from the axis of rotation known as the light
cylinder. Obviously, a poloidal magnetic field discontinuity must
inevitably form beyond the light cylinder (as is shown in the
original Goldreich-Julian cartoon), but that is not too important
since the poloidal magnetic field drops very fast with distance
from the light cylinder.

What is much more important is an equatorial discontinuity in the
azimuthal magnetic field, i.e. the existence of an equatorial {\em
poloidal} current sheet. The first who theorized this
configuration was \citet{1990SvAL...16...16L}, and this led to a
very important school of thought that the acceleration of the
pulsar wind may in fact take place in the equatorial poloidal
current sheet \citep{2001ApJ...547..437L}.
However, his original argument
for the existence of such a current sheet was based on a study of
an artificial solution up to (and {\em not} beyond) the light
cylinder consisting of field lines that close some distance inside
the light cylinder, and field lines that reach the light cylinder.
Obviously, at that time it was completely uncertain what the
latter field lines would do beyond the light cylinder, and unless
a global solution was obtained, no one was in a position to know
with confidence whether such an equatorial {\em poloidal} current
sheet existed or not.

That solution appeared 30 years later (CKF)
and was for several years met with caution
and disbelief \citep{2003PThPh.109..619O}.
Current sheets are usually unstable in
nature, so the equatorial current sheet that CKF obtained was
thought to be artificial. A few authors tried to refute our
discovery proposing various other solutions without current
sheets, but all such efforts did not seem to represent real pulsar
magnetospheres \citep{2006ApJ...652.1494L,
2010mfca.book.....B}. Only after
\citet{2006ApJ...648L..51S} reproduced our solution as the final
steady-state (after twenty rotations) of a
time-dependent numerical evolution did the community begin to pay
attention to the possibility that the equatorial poloidal current
sheet discovered by CKF may be an integral and most important part
of the pulsar magnetosphere. Spitkovky's numerical results have since then
been confirmed by several others \citep{2006MNRAS.368L..30M,
2009A&A...496..495K, 2013MNRAS.435L...1T}.

However difficult the numerical problems associated with
the equatorial current sheet may be (see previous section), we cannot ignore it. 
An important point that we tried to
emphasize during a series of 3 short papers that we wrote many years ago
\citep{2007A&A...475..639C, 2007A&A...472..219C, 2007A&A...466..301C}
is that one cannot say anything about
pulsars without taking proper account of the equatorial current
sheet. As we argued in the previous section, pulsars do not shoot
energetic beams on their own. Pulsar magnetospheres are global
structures that are characterized by what happens both on the
surface of the star (the magnetic field structure, the rotation
rate, and the availability or non-availability of charge carriers
there), and at the equatorial current sheet that develops way out
in the magnetosphere, {\em and not} by the boundary conditions at
`infinity' (as long as the system is able to establish super
Alfv\`{e}nic radiation conditions (${\bf E}\times {\bf
B}/B^2\rightarrow \hat{\bf r}$ at large distances). In particular,
if the physical conditions in the current sheet preclude any
dissipation there, the current sheet establishes itself as a
contact discontinuity with ${\bf B}$ parallel to it above and
below it. This was the case for CKF and all subsequent ideal MHD
simulations. Physically, this implies that the magnetosphere is
able to supply the electric charge carriers needed to support the
electric charge density and electric current density required by
the ideal MHD solution. As long as the required charges are freely
supplied, there is no `residual' electric field along the current
sheet that would lead to particle acceleration along the current
sheet and/or reconnection.

At this point, we would like to describe briefly what takes place
along the equatorial current sheet. As we said, in a contact
discontinuity the magnetic field is parallel to it above (let as
denote it with ${\bf B}_{\parallel {\rm CS}}$), reverses direction inside
the current sheet, and becomes equal in magnitude and antiparallel
below. At the same time, the electric field is perpendicular to it
above the current sheet (let us denote it with ${\bf E}_{\perp {\rm CS}}$),
reverses direction and points in the opposite perpendicular
direction below, thus the current sheet is obviously charged with
surface electric charge density
\begin{equation}
\sigma=\frac{E_{\perp {\rm CS}}}{2\pi}\ .
\end{equation}
Notice that $E_{\perp {\rm CS}}=(r/r_{\rm LC})B_r$ above the current sheet.
In this picture, there is no `residual' electric field {\em along}
the current sheet.
Such a field component (let us denote it with ${\bf E}_{\parallel {\rm CS}}$)
would appear only if the magnetosphere
above and below the current sheet is {\em driven} to converge
towards it with convergence velocity
\begin{equation}
v_{\rm converge}\equiv \frac{{\bf E}_{\parallel {\rm CS}}\times {\bf B}_{\parallel {\rm CS}}}{B^2}c\ .
\end{equation}
Notice that ${\bf E}_{\parallel {\rm CS}}$ lies along the direction of the
electric current in the current sheet, hence it is perpendicular
to ${\bf B}_{\parallel {\rm CS}}$, and continuous (in the same direction)
above and below the current sheet (${\bf E}_{\parallel {\rm CS}}$
becomes radial, $\parallel \hat{\bf r}$, in the current
sheet midplane).
This is often called the reconnection electric field, and in the ideal MHD
region right above and below the current sheet, 
$E_{\parallel {\rm CS}}=(r/r_{\rm LC})B_z$. 
Why this is nonzero depends of course
on the global structure of the magnetosphere. On the other hand,
this is what dictates the dissipation rate at the current sheet,
which is obviously related to the microscopic conditions in the
current sheet. So the `chicken and egg' question is:
\begin{quotation}
{\em What drives the dissipation at the current sheet?}
\end{quotation}
Is it the global magnetospheric structure that redirects electromagnetic
Poynting flux towards the equator, or the local reconnection and
plasma acceleration that dissipates energy in the current sheet?

In a steady state, a balance will be achieved between the Poynting
flux ${\bf E}_{\parallel {\rm CS}}\times {\bf B}_{\parallel {\rm CS}}$ entering the
current sheet and the effective dissipation rate $\eta
E_{\parallel {\rm CS}}^2$ in the current sheet (we have defined here an
effective dissipation parameter $\eta$ such that
$J_{\rm CS}=E_{\parallel {\rm CS}}/\eta$). Thus, the residual electric field
in the current sheet is equal to
\begin{equation}
E_{\parallel {\rm CS}}=\frac{2\eta}{lc}B_{\parallel {\rm CS}}\equiv \frac{2}{{\cal R}_{\rm m}}B_{\parallel {\rm CS}} ,
\label{dissipation}
\end{equation}
where, $l$ is the thickness of the current sheet, and ${\cal
R}_{\rm m}$ is the associated magnetic Reynolds number (or
Lundquist number) based on $l$ and the speed of light.
Unfortunately, this does not answer our question what determined
the dissipation rate. As we will see next, localized
numerical simulations alone can only yield a partial answer to this
question.

The magnetosphere can in principle establish a
CKF-type equilibrium with a dissipationless current sheet, where
all magnetic field lines that cross the light cylinder open up to
infinity. In practice, localized numerical
simulations have shown that, even without external driving, 
such a current sheet becomes unstable to relativistic
Petschek reconnection \citep{2013ApJ...771...54S}.
A finite reconnection speed $v_{\rm converge}\sim 0.1 c$ 
is established which corresponds to the maximum possible
inflow speed for that particular type of reconnection, namely
$v_{\rm maxPetschek}\approx (\pi/8\ln{\cal R}_{\rm
m})c$ \citep[i.e. the reconnection electric
field is practically independent of the local conditions in the
current sheet; see e.g.][]{1985ssmf.conf..121C}. 
This is a local property of the current sheet, and in that case a global
magnetospheric configuration establishes itself such that $v_{\rm
converge}^2+v_{\parallel {\rm CS}}^2\approx c^2$, or equivalently
\begin{equation}
B_r/B_z \sim 9.95
\label{PetschekCS}
\end{equation}
right above and below the equatorial current sheet. Here,
$v_{\parallel {\rm CS}}\equiv (E_{\perp {\rm CS}}/B_{\parallel {\rm CS}})c$\ \ 
is defined as the
drift velocity parallel to the current sheet just above and below it.
As estimated in Eq.~(\ref{PetschekCS}), the entry angle 
of magnetic field lines into the equatorial current sheet
is very small, hence only a small amount of magnetospheric
Poynting flux enters and is dissipated into the current sheet.

Another way to understand this configuration
is to consider the analogy with an
infinite infinitely thin plane conductor threaded by a uniform
current distribution. In that case, a topology of uniform
and opposite direction magnetic field $B_\parallel$ forms above
and below the plane. The source of the field is the current sheet
electric current. If we remove all external power sources and
consider that the sheet is a non-ideal conductor with effective
dissipation
$\eta$, then the current will begin to die out, the current sheet will
absorb electromagnetic energy from above and below, and the
electric current will continue to flow as long as the magnetic
field reservoir is not depleted (i.e. as long as there exist a magnetic field
above and below the plane). In that case, 
the driver of magnetospheric dissipation is the local
dissipation rate in the equatorial current sheet.

We are a little concerned here that
Petschek-type reconnection cannot proceed for too long. As is
seen in numerical simulations, closed magnetic field islands form
along the current sheet, and when the whole current sheet layer is
filled with such islands, the process of tearing and island
formation will not proceed any longer and that particular type of
reconnection will saturate. This does not preclude of course the
occurence of random strong localized energetic reconnection events
along the current sheet which may yield powerful flashes of
high-energy and/or radio emission \citep{2013ApJ...770..147C,
2014ApJ...782..104C}. If our concern is justified then,
however exciting the results of localized current sheet PIC
numerical simulations may be, they do not represent what really
goes on in steady state in the pulsar magnetosphere. In fact the
latest `ab initio' global magnetospheric simulations yield values
of $B_r/B_\perp$ much smaller than 9.95 (on the order of
unity near the light cylinder).
%
%
%

%
%
%
On the other hand, there is strong indirect
evidence that strong dissipation {\em does} take place in the current
sheet \citep[e.g.][]{2015arXiv151101785C,
2014ApJ...793...97K,
2014arXiv1402.1520G}. So, how is the current sheet forced and how
much? PIC numerical simulations do help, provided we have a clearer
idea of how to interpret them. So, let us now consider an
important feature of the equatorial current sheet, the so called Y-point, that may hold
the answer to its vertical forcing.

\section{The Y-point}

Current sheets have a thickness comparable to the gyro-radius
of the particles that support their electric current. It
is well known that these particles follow meandering (snake)
orbits and if a reconnecting electric field exists along the
current sheet (i.e. if there is dissipation), these particles are
accelerated along the direction of $E_{\parallel {\rm CS}}$ following Spiser
orbits, i.e. meandering orbits that become narrower accross and
more stretched along the direction of acceleration. In the
particular case of the pulsar equatorial current sheet this is
true only for the main charge carrier of the current sheet, the
one with the sign of charge of the charged sheet. This charge
carrier is accelerated outwards. The other charge carrier is
accelerated inwards, but in doing so it meanders all the way to
the outskirts (the surfaces) of the current sheet where the main
${\bf E}\times {\bf B}$ direction is still outwards. So, the
meandering of the other charge carrier is `messy', with
excursions outside the current sheet. Nevertheless, on average, these too are
accelerated inwards. However, it could be that they cannot survive
for much time along the current sheet. This has already been seen
in numerical simulations \citep{2007A&A...472..219C,
2015ApJ...801L..19P}, and remains to be
confirmed with further studies.

The particle gyro-radius is determined by the particle kinetic
energy perpendicular to the magnetic field. In the pulsar
magnetosphere, the perpendicular kinetic energy is quickly
radiated away through curvature radiation, and if most of the
kinetic energy is radiated away, then the particle gyro-radius
becomes smaller and smaller, and the current sheet narrower and
narrower. In the limit of very small gyro-radius the current
sheet can still be held from collapsing by its own electrostatic
repulsion \citep[see][for details]{2014ApJ...781...46C}.
In other words, the current sheet
will never completely collapse because it is held by its own electrostatic
repulsion. This is true unless we sit right at the Y-point.

As we have argued before, the poloidal magnetic field, thus also
the poloidal electric field and the electric charge density all
three vanish right outside (but not inside) the Y-point. It
becomes then extremelly problematic to support a current sheet
right outside the Y-point, unless of course the external magnetic
pressure (i.e. $B_\phi$) also vanishes at that point! This simple
result has very strong implications for the global structure of
the magnetosphere. Firstly, this is equivalent to the vanishing of
the electric current along the separatrix between the corotating
dead zone and the open magnetosphere. There is no current sheet
that returns to the star. The equatorial current sheet does not
close the electric circuit onto the star. And if we do not allow
for electric currents to flow through the Y-point, then the old
CKF-type solution cannot be valid anymore, and the magnetosphere
must find a new global equilibrium that is {\em different}
from the CKF solution. In fact, we expect this equilibrium to be strongly
dissipative in the current sheet for a simple reason: CKF argued
that their solution is the only possible solution for a dissipantionless
ideal magnetosphere. The new solution is still ideal everywhere,
but very different in the current sheet, thus also strongly dissipative there.
The equatorial current sheet
must be gradually replenished by the magnetospheric electric
charges that both support it vertically and also form the electric
current that threads it. In doing so, Poynting flux enters the
current sheet, and this energy flux must find a way to dissipate
in the current sheet through reconnection and particle
acceleration (with ensuing emission of radiation). 
We have thus
found the origin of the magnetospheric forcing of the current
sheet, namely the inability of the electric current in the current
sheet to cross the Y-point.

The magnetospheric Y-point is indeed a problematic place and
several authors have studied it in detail in the past. It cannot
approach too close to the light cylinder because in that case the
corotation velocity at the tip of the dead zone would approach the
speed of light and the magnetic and electric fields right inside
it would diverge \citep{2003ApJ...598..446U}. At the same time, it
must be in a position to support a sudden drop in its charge density. We
will now present why we believe it is hard in practice for the Y-point to also support
a continuous electric current through it. Before we proceed, let us here define as
the `right' type of particles, those that, when they move outwards,
they support the magnetospheric electric current, and `wrong' the
other type. For example, in the case of the equatorial current sheet in the
aligned rotator ($\Omega$ parallel to the magnetic moment of the
star), the positrons/protons/ions are of the `right' type, and
electrons of the `wrong' one.

In order to support at the same time a continuous current, a
sudden density drop across, and a zero charge density just
outside the Y-point, we must have two populations of particles
that satisfy both
\begin{equation}
\rho_{\rm wrong}+\rho_{\rm right}=0\ ,\ \mbox{and}
\end{equation}
\begin{equation}
\rho_{\rm wrong}v_{\rm wrong}+\rho_{\rm right}v_{\rm right}=J_{\rm Y}\ .
\label{JY}
\end{equation}
However small the amount of dissipation in the equatorial
current sheet \citep[i.e. even if we just have spontaneous
relativistic Petschek-type reconnection as shown
in the localized numerical simulations of][]{2013ApJ...771...54S},
then its electric field accelerates particles of the wrong type
inwards. Therefore $v_{\rm wrong}$ is inward and particles of the
wrong sign must follow {\em inward} meandering
(snake) orbits traveling from outside the light cylinder into the
Y-point. Unfortunately, this is not
allowed topologically in a 3D magnetosphere (i.e. oblique rotator)
for the following reason: outward moving particles of the right
type do support a trailing (retarded) spiral current sheet (in the
shape of a spinning balarina skirt). Outside the light cylinder,
particles cannot move inwards along such a 3D current sheet, thus
$\rho_{\rm wrong}$ must vanish outside the light cylinder. According
to eq.~(\ref{JY}) $J_{\rm Y}$ must also vanish there.

We are thus left with one possibility, namely
a strongly dissipative equatorial current sheet with no electric
current at the Y-point. This is the new standard magnetosphere of 
\citet{2014ApJ...781...46C}. We plan to investigate this solution
in greater detail in the future. Meanwhile, let us go back to the
old standard magnetosphere (CKF) and discuss certain of its
features that have not been properly emphasized in the literature.

\section{The crossings of the null surface}

In the old standard magnetosphere,
there is a line (in 3D a surface) along which the Goldreich-Julian
space charge density
vanishes. This is the so called null line (or null surface in 3D).
%
%
%
%
In relation to the null line, there are four
types of field lines (see Fig.~2):
\begin{enumerate}
\item Open field lines that do not cross the null line. These
originate on the stellar polar cap around the magnetic axis
(which in axisymmetry coincides with the axis of rotation), and carry
the main part of the polar cap poloidal electric current. \item
Closed field lines. All of them do cross the null line. These
originate at low stellar latitudes up to the polar cap, and do not
carry any poloidal electric current. \item Open field lines that
cross the null line once. These originate at the outskirts of the
polar cap, and carry some fraction of the return poloidal electric
current. \item Open field lines that cross the null line twice.
These originate somewhere inside the polar cap, and carry a very
small amount of electric current of the same polarity as the main
part of the polar cap.
\end{enumerate}
Let us now consider each field line category and propose how it
may be filled with charged particles. Notice that the source of
particles may very well be a stellar reservoir of protons (ions)
and electrons, and does not need to resort to pair formation in
magnetospheric gaps.

\begin{figure}
\centering
\includegraphics[trim={0cm 2cm 0cm 4cm}, width=\textwidth]{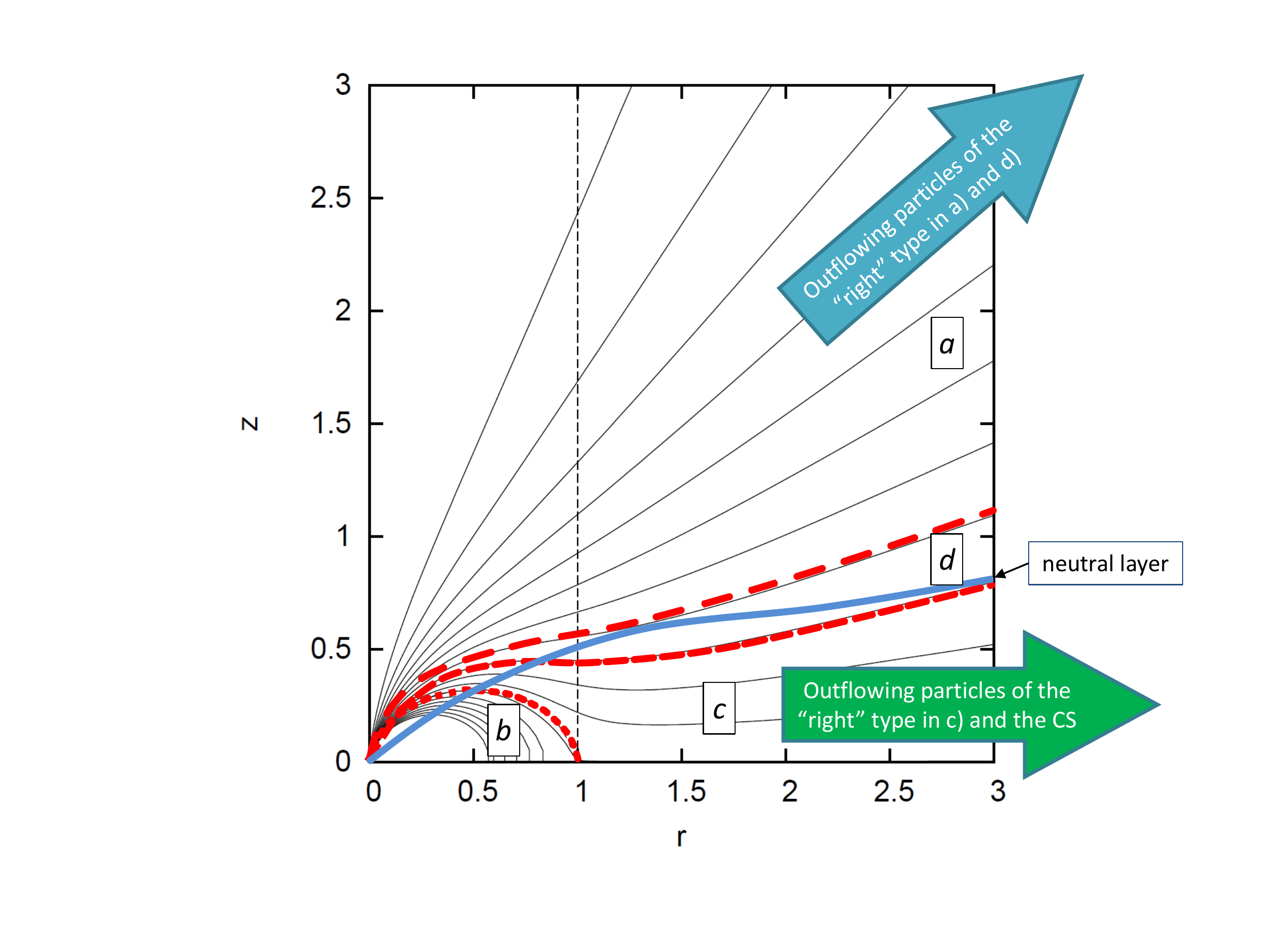}
\caption{The crossings of the null surface in the
standard ideal force-free axisymmetric CKF magnetosphere.
Vertical dashed line: the light cylinder. Blue line: null line (the neutral layer).
The three red lines separate the four types of field lines (see text for
details).
Type a: above red long dashed line. Type b: below red dotted line. Type c:
between red dotted and short dashed lines. Type d: between red short and long 
dashed lines. The red short dashed line is the line of zero poloidal electric current
$J_{\rm CKF}$ (the electric current null line) that separates the magnetosphere
into two regions of outgoing and ingoing electric currents respectively. 
For example, in the case of aligned pulsars, the `right' type of particles 
are outflowing electrons in regions a) and d) (large blue arrow), and outflowing protons or
positrons in region c) and in the equatorial current sheet (CS; large green arrow). The electric charge
and electric current null lines become one asymptotically (as $r\rightarrow\infty$).}
\label{Fig2}
\end{figure}

The first type of field lines can very well be filled {\em only
with one} type of particles, the one that we call `right'. In this
part of the magnetosphere, the `right' type can perfectly
accommodate both the charge and electric current, namely
$\rho=\rho_{\rm GJ}$ everywhere (in fact almost everywhere as we
will see next), and $J_{\rm CKF}=\rho_{\rm GJ} v_{\rm poloidal}$. Following
our definition of $J_{\rm CKF}$ and $J_{\rm GJ}$ in \S~1.2, one
can easily check that
\begin{equation}
J_{\rm CKF}\ll J_{\rm GJ}\ ,
\end{equation}
near the stellar surface, and $J_{\rm CKF}$ flows in the same
direction as $J_{\rm GJ}$, hence only one type of charge carrier
is sufficient to accomodate both the charge and electric
current densities by adjusting the value of $v_{\rm poloidal}\ll
c$, and
\begin{equation}
J_{\rm CKF}\rightarrow J_{\rm GJ}
\label{Jasymptotic}
\end{equation}
at large distances, where $v_{\rm poloidal}\rightarrow c$. There is
nothing terribly interesting to discuss in this region. There is a
slight complication at the outskirts of this region where field
lines at some point approach close to the null line, and at those
positions the adjustment of $v_{\rm poloidal}\equiv J_{\rm
CKF}/\rho_{\rm GJ}$ does not work because the flow would become
superluminal, but we will discuss this in relation to the field
lines of the fourth type as defined above.

The second type of field lines does not have any problem
whatsoever to be filled with two types of particles, positive and
negative, that simply corotate with the star and thus form the so
called dead zone. Nothing is moving in the poloidal direction, and
the poloidal electric current is zero. The particles are held in
corotation by microscopic electric forces that are outside the
realm of numerical ideal MHD analysis.

The third type of field lines is much more interesting. These
field lines start with the `wrong' type of charge carriers on the
stellar surface, and switch to the `right' type beyond the null
surface. Asymptotically, Eq.~(\ref{Jasymptotic}) still holds
where $v_{\rm poloidal}\rightarrow c$. So these field lines need
to {\em at least} carry just the Goldreich-Julian charge density
at large distances. These particles of the `right' type can very
well originate on the stellar surface and support the electric
current all the way to the star. What needs to be done is to also
pull the right amount of `wrong' type particles from the stellar
surface in order just to fill the region from the stellar surface
up to the null line and some small distance beyond. In a configuration
with the {\em minimal possible number of particles},
the innermost part of the third
type of field lines contains a corotating region of the `wrong'
type of particles with density
\begin{equation}
\rho_{\rm corotating}\approx\rho_{\rm GJ}-\frac{J_{\rm CKF}}{c}\
.
\end{equation}
Obviously, since $\rho_{\rm GJ}$ becomes zero some distance
downstream along these lines, the corotating region of the `wrong'
type particles must extend some distance further downstream from
the null line. We suspect that the region of corotating charge
never extends beyond the light cylinder (where corotation is
obviously not supported), and this is helped by the fact that $J_{\rm
CKF}\rightarrow 0$ at the boundary between regions three and four.

The fourth type of field lines is the most interesting of all.
These carry an electric current which can very well be
accommodated only by charge carriers of the `right' type (which is
the same type as the `right' type for region one) since 
$J_{\rm CKF}\rightarrow J_{\rm GJ}$
at large distances, but consists of three regions of electric
charge, of the `right' type near the star and at very large
distances, and of the `wrong' type in between. The most economical
obvious way to satisfy all the requirements in this region is to
increase the number density of particles of the `right' type, and
add a population of outflowing particles of the `wrong' type which
will adjust their poloidal velocities in order to build the in
between region of the `wrong' type Goldreich-Julian density. This
configuration will extend also to the outskirts of region one in
order to avoid superluminal velocities for particles of the
`right' type.

In any case, the poloidal speed of the charge carriers of the
`wrong' type will be much smaller than the speed of the charge
carriers of the `right' type. Thus, in regions three and four
(extended also to the outskirts of region one), it is very natural
to expect a configuration where a population of relativistic
$v\approx c$ particles of one type flows through a large
population of almost stationary particles of the other type. Such
a configuration may naturally yield Cherenkov type instabilities
that may be related to the pulsar radio emission, as discussed e.g. in
\citet{1999ApJ...512..804L}. This scenario certainly merits 
further investigation.

\section{Epilogue}

We have constructed a very economical pulsar magnetosphere
consisting of large corotating regions filled with particles of one 
type of charge, threaded by streams of fast outflowing particles
of the other type of charge. It is thus possible to
fill the magnetosphere with a minimum amount of charge carriers by
adjusting their speed and concentration that does not require
continuous pair formation in the magnetosphere. In fact, particles
of both negative and positive charge can naturally be found in
the neutron star atmosphere with no need to resort to pair
formation. Nature manages its resources wisely,
so, instead of spending energy to produce a
multitude of pairs, it can fill the magnetosphere with
protons and electrons provided freely by the star.
In that case, aligned rotators will obviously
differ from counter-aligned ones. In particular,
the equatorial current sheet of aligned pulsars will contain protons 
which will produce a markedly different high energy signature
than that of counter-aligned pulsars which contains electrons. Moreover,
aligned pulsars will contain large corotating regions
filled with electrons which may also produce a 
radio signal different from that of counter-aligned pulsars.
Future observations may indeed be able to discriminate
between aligned and counter-aligned pulsars.

We have also emphasized certain restrictions in the equatorial
current sheet that may prevent it from extending all the way to
the stellar surface. The latter dramatically modify
the structure of the pulsar magnetosphere away from the standard
ideal MHD solution. Obtaining the new standard solution required the
investigation of several magnetospheric issues in detail that
cannot be reached in modern state-of-the-art 3D numerical
simulations. The most recent numerical simulations yield vague hints
that the magnetospheric structure is closer to the new standard
solution, but are not yet in a position to reveal its details.
We believe that it is too early to run global `ab initio'
magnetospheric PIC simulations since we need to understand
in greater depth how the local conditions around the current sheet
affect the global magnetospheric structure.
Therefore, what we propose instead
is to focus future PIC simulations in the study of the equatorial current sheet
and the Y-point, and combine them with ideal MHD/FFE simulations of the global
magnetosphere.

Our numerical simulations were performed using the resources of 
the NRNU MEPhI High-Performance Computing Center.

\newpage
\bibliographystyle{jpp}

\bibliography{TheElectrostaticPCS}

\end{document}